\documentclass[journal=langd5,manuscript=article,layout=twocolumn]{achemso}

\usepackage{gensymb} % to be able to write degrees (¡) with the command \degree

\newcommand*{\sometext}{We study the spatio-temporal dynamics of water uptake by capillary condensation from unsaturated vapor in mesoporous silicon layers (pore radius $r_\mathrm{p} \simeq 2$ nm), taking advantage of the local changes in optical reflectance as a function of water saturation. Our experiments elucidate two qualitatively different regimes as a function of the imposed external vapor pressure: for low saturations, equilibration occurs via a diffusion-like process; for high saturations, an imbibition-like wetting front results in fast equilibration towards a fully saturated sample. We show that the imbibition dynamics can be described by a modified Lucas-Washburn equation that takes into account the liquid stresses implied by Kelvin equation.}

\let\oldmaketitle\maketitle
\let\maketitle\relax

\author{Olivier Vincent}
\email{olivier.vincent@cornell.edu}
\author{Bastien Marguet}

\author{Abraham D. Stroock}
\email{ads10@cornell.edu}
\affiliation[]
{Robert Frederick Smith School of Chemical and Biomolecular Engineering, Cornell University, 120 Olin Hall, Ithaca NY 14853}
\title[Running Title]
  {Imbibition triggered by capillary condensation in nanopores}


\begin{document}

\twocolumn[
\begin{@twocolumnfalse}
\oldmaketitle
\begin{abstract}
\sometext
\end{abstract}
\end{@twocolumnfalse}
]

%\begin{tocentry}
%
%\includegraphics{Figures/GraphicalAbstract}
%
%\end{tocentry}



\section{Introduction}

The interaction of porous media with liquids and vapors occurs in many contexts. In nature, water is ubiquitous within soils, rocks, or the atmosphere, and its behavior in pores is fundamental for the hydration and of plants and their vascular flows\cite{Stroock2014}, the stability of sand terrains \cite{Kudrolli2008}, and the formation of clouds by condensation on atmospheric particles \cite{Marcolli2014}. In technology, the permeation of liquids through porous media occurs in a diversity of applications, from printing ink on paper to oil recovery \cite{Alava2004}. As a result, the dynamics of invasion of a pore space when put in contact with liquid, i.e. imbibition, has been the subject of many studies, from the early work of Lucas and Washburn \cite{Lucas1918,Washburn1921} to recent work on nanoporous media \cite{Acquaroli2011,Gruener2012}. The observed dynamics of invasion show excellent agreement with the classical Lucas-Washburn equations, even in pores approaching the molecular size \cite{Huber2015,Vincent2016}.

Nanopores, however, can fill with liquid even if in contact with the vapor phase only. This phenomenon, known as capillary condensation, occurs for any pore size but persists for vapor pressures significantly below saturation only for diameters in the nanometer range \cite{Charlaix2010}. Capillary condensation plays a role in may areas of science and technology such as the cohesion and friction of granular materials \cite{Bocquet1998}, or the dynamics of hydrocarbons in subsurface reservoirs \cite{Barsotti2016}. Adsorption by capillary condensation, and the reverse process of desorption are still subject of active research, with outstanding questions related to the collective effects induced by disorder in the pore network \cite{Wallacher2004,Aubry2014} or to deviations from macroscopic thermodynamics and dynamics at the nanoscale \cite{Neimark2003,Vincent2016}.

The experimental study of both imbibition and capillary condensation at the nanoscale is challenging due to the difficulty of accessing the local filling state of the porous medium: often only the global mass uptake can be recorded \cite{Valiullin2006,Gruener2009}. Spatially resolved information can however be obtained by more sophisticated techniques such as interferometry \cite{Acquaroli2011}, neutron radiography \cite{Gruener2012} or capacitance measurements \cite{Kiepsch2016}. Here, we use the local changes in reflectance of quasi two-dimensional, laterally connected porous silicon layers to study the dynamics of water uptake as a function of the imposed water vapor pressure (relative humidity) around the medium. We elucidate a qualitative transition as a function of humidity from a diffusion-like regime to an imbibition-like regime. For the imbibition-like regime, we show that a modified Lucas-Washburn equation that takes into account the effective capillary pressure predicted from Kelvin-Laplace equation explains our observations quantitatively.

\begin{figure*}
  	\begin{center}
	\includegraphics[scale=1.1]{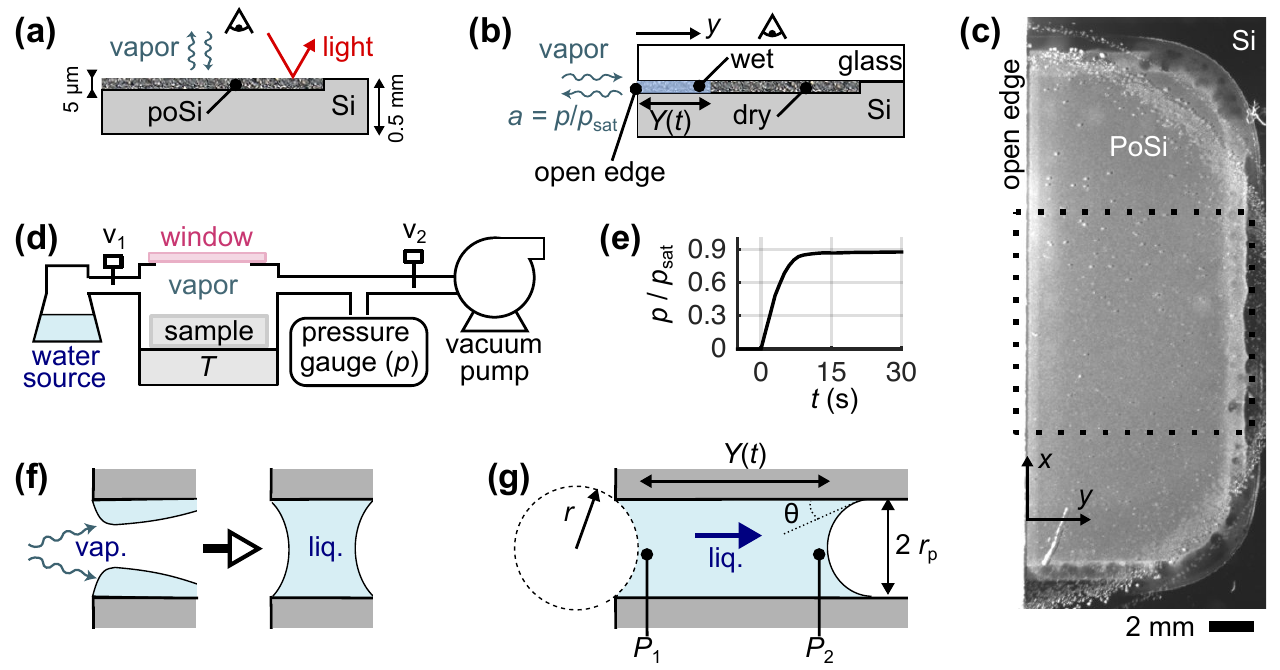}
 	 \caption{
	 \small
	 Side-view sketch of the sample before (a) and after (b) bonding to glass. 	(c) Top-view photograph of the sample. The dashed square shows the portion of the sample displayed in Figure \ref{fig:ImbibitionDynamics}a. (d) Sketch of the vacuum system used to control the vapor pressure and the temperature. (e) Vapor pressure signal during an experimental run, showing a typical early transient. (f) Schematic of the process of capillary condensation of vapor into liquid in a pore. (g) Pore-scale representation of the dynamics of liquid invasion following condensation at high vapor pressure, driven by the competition between the capillary pressures $P_1$ (determined by Kelvin equilibrium: Equation \ref{eq:Kelvin}) and $P_2$ (determined by the geometry of the pore and the equilibrium contact angle of the liquid on the pore wall: Equation \ref{eq:PsiC}).
	 	 }
  	\label{fig:Intro}
	\end{center}
\end{figure*}

\section{Methods \label{sec:Methods}}

We formed a $5 \, \mathrm{\mu m}$ thick layer of mesoporous silicon (PoSi) on top of a p-type silicon (Si) wafer of $\langle 111 \rangle$ crystal orientation and $1\--10 \, \mathrm{\Omega.cm}$ resistivity, by anodization in a mixture of 1:1 $49\%$ hydrofluoric acid:pure ethanol at $20 \, \mathrm{mA/cm^2}$ current density for 5 minutes. We then used thermal oxidation at $700\degree$C  in pure oxygen for $30$ seconds to increase hydrophilicity and stabilize the porous layer. From previous studies we expect a laterally connected pore structure with porosity, $\phi=0.45$ and a typical pore radius, $r_\mathrm{p}=1.5 \-- 2$ nm \cite{Vincent2014,Vincent2016}.

The sample resulting from these processing steps is sketched in Figure \ref{fig:Intro}a. Such a sample could equilibrate quickly with the environment due to the open top surface and to the small thickness of the porous silicon layer ($\sim \mathrm{\mu m}$). For that reason, we measured the reflectance isotherms (see below) at this step in the fabrication process. We then sealed the top surface to glass by anodic bonding (Figure \ref{fig:Intro}b), so that the sample could exchange vapor only at its open edge; this change resulted in much slower equilibration dynamics in the $y$ direction due to the large lateral dimensions of the porous layer ($\sim \mathrm{cm}$). Figure \ref{fig:Intro}c presents a top-view photograph of a sample as in Figure \ref{fig:Intro}b.

We placed the sample in a vacuum chamber equipped with an optical window, thermostated at $T=15 \degree$C (Figure \ref{fig:Intro}d). Pure water vapor, obtained by evaporation from a degassed liquid water source, flowed through the chamber towards the vacuum pump, and we measured its pressure $p$ using a pressure gauge connected to the chamber. By adjusting the relative opening of two needle valves $\mathrm{v_1}$ and $\mathrm{v_2}$, we could set the pressure between $p=0$ and $p=p_\mathrm{sat}$. All capillary condensation experiments started with a dry sample equilibrated for at least $24$ hours at $p=0$ ($\mathrm{v1}$ closed). Opening v1 at a time defined as $t=0$ allowed us to impose a steady, non-zero vapor pressure $p$ after a transient time of typically $\simeq 10$ seconds (Figure \ref{fig:Intro}e). From the error of the pressure gauge reading as well as the measured fluctuations of $T$ and $p$ during the experiments, we estimated an uncertainty on the activity, $a = p/p_\mathrm{sat}$ of $\pm 0.005$.

We recorded time-lapse image sequences of the sample, and analyzed its response to changes in vapor pressure by extracting the local relative change in reflectance $\Delta I / I$ under constant, white-light, diffuse illumination. From the images, we calculated $(\Delta I / I)_i = (g_i^\mathrm{ref} - g_i)/g_i^\mathrm{ref}$ where $g_i$ was the grayscale value of the $i^\mathrm{th}$ pixel in the the image ($g_i \in[0\--255]$) and $g_i^\mathrm{ref}$ was its average value for $t<0$, corresponding to the dry sample. Water uptake produced a darkening of the image, resulting in $(\Delta I / I)_i > 0$. For the measurement of the reflectance isotherm, we evaluated the average change in intensity, $\langle \Delta I / I \rangle = \frac{1}{N} \sum_i (\Delta I / I)_i$ over the whole region of interest shown in Figure \ref{fig:Intro}c ($\langle \Delta I / I \rangle_{xy}$), while for the extraction of invasion dynamics propagating along $y$, averaging was done along the $x$ direction only ($\langle \Delta I / I \rangle_x (y)$, axes shown in Figure \ref{fig:Intro}c).

\section{Theory}

\paragraph{Kelvin Equation}
We consider liquid-vapor equilibrium at temperature $T$ with liquid at pressure $P$ and vapor at pressure $p$. It is convenient to define the activity, $a$, of the vapor, which compares the vapor pressure to its saturation value $p_\mathrm{sat}(T)$,
\begin{equation}
a = \frac{p}{p_\mathrm{sat}}.
\label{eq:Activity}
\end{equation}
Note that the quantity $a\times100$ is also the percent relative humidity of the vapor $\mathrm{[\%RH]}$.
Bulk equilibrium between liquid and vapor imposes $P = p =  p_\mathrm{sat}(T)$. In a porous medium, however, the liquid and vapor phase can coexist at different pressures due to the capillary pressure associated with the curved liquid-vapor meniscus. From Laplace equation $P - p = \sigma \mathcal{C}$, where $\mathcal{C} \, \mathrm{[m^{-1}]}$ is the curvature of the meniscus and $\sigma \, \mathrm{[N/m]}$ is the surface tension of the liquid. The condition for equilibrium, even with $P \neq p$, is the equality of the chemical potentials $\mu_\mathrm{liq}$ and $\mu_\mathrm{vap}$ in the liquid and vapor phases. Integration of the isothermal Gibbs-Duhem equation, taking $P=p=p_\mathrm{sat}$ (chemical potential $\mu_0$) as a reference state,  yields $\mu_\mathrm{liq} = \mu_0 + v_\mathrm{m} (P - p_\mathrm{sat})$ and $\mu_\mathrm{vap} = \mu_0 + RT \ln \left( a \right)$, where $v_\mathrm{m}$ is the molar volume of the liquid, and $R$ is the universal gas constant. The expression of $\mu_\mathrm{vap}$ uses the assumption that the vapor behaves as a perfect gas, while that of $\mu_\mathrm{liq}$ is obtained assuming an incompressible liquid ($v_\mathrm{m}$ independent of pressure, which results in negligible errors for water at room temperature \cite{Vincent2016}). Eventually, the equilibrium condition through the equality of chemical potential requires
\begin{equation}
P = p_\mathrm{sat} + \Psi (a)
\label{eq:Kelvin}
\end{equation}
where we have defined the vapor water potential
\begin{equation}
\Psi(a) = \frac{RT}{v_\mathrm{m}} \ln \left(  a \right).
\label{eq:Psi}
\end{equation}
Equation \ref{eq:Kelvin} is known as Kelvin equation, and is based on macroscopic thermodynamic considerations and descriptions of the pore fluid. Kelvin equation has however been shown to accurately describe liquid capillary stresses due to subsaturated liquid-vapor equilibrium, even in nanoscale confinements \cite{Fisher1980,Gor2015,Vincent2016}.
\paragraph{Liquid invasion dynamics}
At high vapor pressure, it becomes thermodynamically favorable for the pores to fill with liquid by the process of capillary condensation \cite{Charlaix2010}. Capillary condensation causes the formation of liquid plugs at the sample edge (Figure \ref{fig:Intro}f), which progressively invade the whole depth of the pores. We assume that the pore filling after capillary condensation proceeds by bulk liquid flow driven by the competition of capillary pressures $P_1$ and $P_2$ as depicted in Figure \ref{fig:Intro}g, and that condensation of vapor at the edge (meniscus 1) provides the liquid necessary to account for the advance of meniscus 2. In other words, we consider that the pore wetting dynamics is slow compared to the dynamics of vapor condensation at the edge \cite{Note-1}, so that meniscus 1 is in local thermodynamic equilibrium with the vapor. As a result, from Equations \ref{eq:Kelvin}-\ref{eq:Psi}, $P_1 = \Psi (a)$, where we have neglected $p_\mathrm{sat}$ compared to $\Psi$. We assume that the curvature of meniscus 2 is defined by the local mechanical equilibrium of the triple line, i.e., the curvature is $\mathcal{C} = - 2 \cos \theta / r_\mathrm{p}$ where $\theta$ is the equilibrium contact angle of the liquid on the solid and $r_\mathrm{p}$ the radius of the pore. From Laplace equation, $P_2 - p = \Psi_\mathrm{c} $ where 
\begin{equation}
\Psi_\mathrm{c} = - \frac{2 \sigma \cos \theta}{r_\mathrm{p}}
\label{eq:PsiC}
\end{equation}
is the intrinsic capillary pressure of the pore. Neglecting the vapor pressure $p$ compared to $P_2$, we have $P_2 = \Psi_\mathrm{c}$ and
\begin{equation}
\Delta P = P_1 - P_2 = \Psi (a) - \Psi_\mathrm{c}.
\label{eq:DeltaP}
\end{equation}
The driving force for the flow is the pressure gradient $\nabla P = \Delta P / Y(t)$, where $Y(t)$ is the location of meniscus 2 (Figure \ref{fig:Intro}g). We describe the flow response to that gradient using Darcy equation $q = \rho_\mathrm{liq} \kappa \nabla P$, where $q \, \mathrm{[kg/(m^2.s)]}$ is the mass flux, $\rho_\mathrm{liq}$ is the density of the liquid, and $\kappa \, \mathrm{[m^2/(Pa.s)]}$ is the permeability of the porous layer. Using the conservation of mass $q = \rho_\mathrm{liq} \phi \,  \mathrm{d}Y / \mathrm{d}t$, the front position $Y(t)$ obeys the differential equation $Y \mathrm{d}Y / \mathrm{d}t = \kappa \Delta P / \phi$. Integration using $Y=0$ at $t=0$ yields
\begin{equation}
\frac{Y^2(t)}{t}= w = \frac{2 \kappa}{\phi} \left( \Psi(a) - \Psi_\mathrm{c} \right)
\label{eq:Washburn}
\end{equation}
where we have used Equation \ref{eq:DeltaP} to express the pressure difference $\Delta P$. When $\Psi (a)=0$ (i.e. $p=p_\mathrm{sat}$ from Equations \ref{eq:Activity} and \ref{eq:Psi}), Equation \ref{eq:Washburn} reduces to the Lucas-Washburn law for the dynamics of imbibition, which describes the non-inertial dynamics of liquid invasion when the sample edge is plunged in bulk liquid \cite{WickingBook2013}. The extra term $\Psi (a)$ accounts for the existence of an additional capillary pressure governed by Kelvin equation when the sample is in contact with vapor instead of liquid (Figure \ref{fig:Intro}g). We will refer to the parameter $w$ as the \emph{imbibition speed coefficient} in the rest of the paper.

%\Oc{- Hypotheses not explicited in the theory section above: 1) liquid in pore behaves the same as bulk liquid for Kelvin equation, 2) Inertial effects are neglected in Lucas-Washburn, 3) Flow is assumed to be quasi steady-state (linear spatial gradient of pressure), which is ok since poroelastic diffusion is fast compared to imbibition, 4) Contact angle is at equilibrium, can be discussed in terms of the low Capillary number of the flow in the pore.}

\paragraph{Vapor invasion dynamics}

At low vapor pressures, no capillary condensation occurs, but vapor can adsorb on the pore walls. The average density $\rho(a)$ inside of the pore depends on the vapor activity, and varies between $\rho=0$ at $a=0$ (evacuated sample) to $\rho=\rho_\mathrm{liq}$, as $a$ approaches $1$ (pore filled with liquid of density $\rho_\mathrm{liq}$ due to capillary condensation). It is thus convenient to define $\alpha (a) = \rho(a) / \rho_\mathrm{liq}$, which takes values between 0 and 1. The dynamics of vapor invasion in the pores is a result of the combined effects of transport through the pores and adsorption on the pore walls, and obeys the conservation of mass $\partial q / \partial y = - \phi \partial \rho / \partial t$. We define a generic transport coefficient $k(a)$ so that $q = - \phi \rho_\mathrm{liq} k(a) \partial a / \partial y$; the conservation of mass then translates into
\begin{equation}
\alpha^\prime \frac{\partial a}{\partial t} = \frac{\partial }{\partial y}\left( k(a) \frac{\partial a}{\partial y}\right),
\label{eq:VaporPoroelasticity}
\end{equation}
where we have defined $\alpha^\prime = \mathrm{d}\alpha / \mathrm{d}a$ the derivative of the adsorption function. Equation \ref{eq:VaporPoroelasticity} is a complex diffusion equation with activity-dependent coefficients, but for
small variations in activity, it reduces to is a simple diffusion equation with an effective diffusivity
\begin{equation}
D = \frac{k}{ \alpha^\prime}.
\label{eq:VaporDiffusivity}
\end{equation}
Equation \ref{eq:VaporDiffusivity} illustrates the competition between transport and adsorption: when the adsorbed mass changes rapidly as a function of vapor pressure ($\alpha^\prime$ large), adsorption on the pore walls act as a sink, slowing down the propagation of the vapor. Equations \ref{eq:VaporPoroelasticity}-\ref{eq:VaporDiffusivity} are very general and will serve as a basis for discussion of the possible mechanisms that are at play during equilibration of the samples with vapor.

The generic transport parameter $k$ incorporates the physics of transport. If transport is dominated by diffusion in the vapor phase, by Fick's law we have
 $q = \phi M D_\mathrm{m} / \tau \times \partial C / \partial y$, where $D_\mathrm{m}$ is the molecular diffusivity, $M$ is the molar mass, and $C=p/(RT)$ is the concentration, assuming that the vapor behaves as an ideal gas, and $\tau$ is the tortuosity of the pore space. The transport coefficient is then $k = D_\mathrm{m} p_\mathrm{sat} v_\mathrm{m} / (\tau RT)$. In nanometer-size pores, the diffusivity can be estimated as $D_\mathrm{m} = (2 r_\mathrm{p} / 3) \sqrt{8RT/(\pi M)}$ from Knudsen diffusion \cite{Gruener2008}, yielding $D_\mathrm{m} \simeq 8 \times 10^{-7} \, \mathrm{m^2/s}$ for water vapor in pores 2nm in radius, corresponding to a transport coefficient $k \simeq 2 \times 10^{-12} \, \mathrm{m^2/s}$. Assuming $\alpha^\prime \sim 1$ for typical adsorption isotherms, we expect $D$ to also be on the order of $10^{-12} \, \mathrm{m^2/s}$ if transport is based on Knudsen diffusion. This value will serve as a reference in our discussion of the diffusion mechanisms below.

%\Oc{- Note:  when no adsorption occurs, the density $\rho$ is simply that of the vapor, and the function $\alpha$ reduces to $\alpha = v_\mathrm{m} p / (RT)$, and its derivative with respect to $a$ is $\alpha^\prime=v_\mathrm{m}p_\mathrm{sat}/(RT)$. Consequently, Equation \ref{eq:VaporPoroelasticity} then simply reduces to the diffusion equation with $D = D_\mathrm{m}$.}

%\Oc{- Since the model above is not that successful at describing the experimental results, should we only put it as Supporting Information?}

%- Data obtained on 2 samples, 3 open edges (from same wafer)

%-
%- As in previous work, we use water potential to describe liquid and vapor in a unified way (or introduce later?)

%- Porosity $\phi = 0.45$ from previous study \cite{Vincent2016}.

\section{Results}

\paragraph{Reflectance isotherm}

\begin{figure}
  	\begin{center}
	\includegraphics[scale=1]{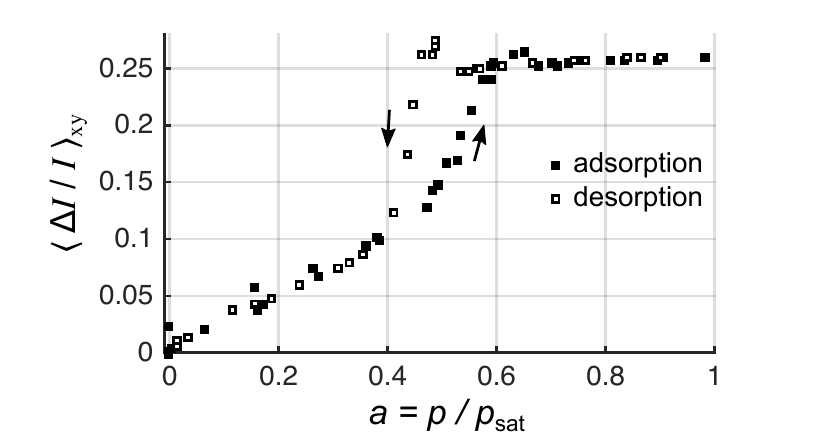}
 	 \caption{
	 \small
	 Reflectance isotherm of the sample measured on the open sample (Figure \ref{fig:Intro}a). $ \langle \Delta I / I \rangle_{xy}$ represents the relative change in average grayscale intensity with respect to the dry sample (see Methods). $\langle \Delta I / I \rangle_{xy} > 0$ corresponds to a darker image and to a higher sample saturation.
	  }
  	\label{fig:Isotherm}
	\end{center}
\end{figure}

We first recorded the static changes in reflectance of the porous silicon layer as a function of the vapor activity $a = p / p_\mathrm{sat}$ with an open sample (not bonded to glass) as sketched in Figure \ref{fig:Intro}a. The resulting \emph{reflectance isotherm} is presented in Figure \ref{fig:Isotherm}. These reflectance isotherms display features characteristic of adsorption and desorption isotherms measured by mass in nanoporous media \cite{Charlaix2010}: a reversible plateau at high relative humidities (here for $a>0.65$), associated with the presence of bulk liquid in all the pores; a reversible branch at low relative humidities (here for $a<0.4$), associated with the presence of adsorbed molecules on the pore walls; and a hysteresis loop at intermediate relative humidities, with desorption occurring at lower pressure than adsorption. Although the nature of the hysteresis loop and its relationship to pore structure are still debated, the adsorption branch is typically interpreted as the result of capillary condensation \cite{Charlaix2010,Huber2015}.

The close resemblance between the reflectance isotherm and mass isotherms for water in porous silicon \cite{Kovacs2009} suggests that there is a direct, close-to-linear relationship between adsorbed mass and reflectance, and thus that studying optical changes is a useful tool to access local concentration changes during dynamic experiments.
Compared to a mass isotherm, we note additional features. In particular, "spikes" stand out above the hysteresis loop at the edge of desorption (for $a$ slightly below $0.5$) and in the later stages of capillary condensation (for $a \simeq 0.6\--0.65$). We interpret the presence of these spikes as a loss of reflectance associated with increased light diffusion in the porous medium due to the formation of large clusters of liquid and vapor of size comparable to the wavelength of light. Such effects have been reported in Vycor porous glass both for light and sound propagation \cite{Page1993,Soprunyuk2003}.

\paragraph{Invasion dynamics}

\begin{figure}
  	\begin{center}
	\includegraphics[scale=1]{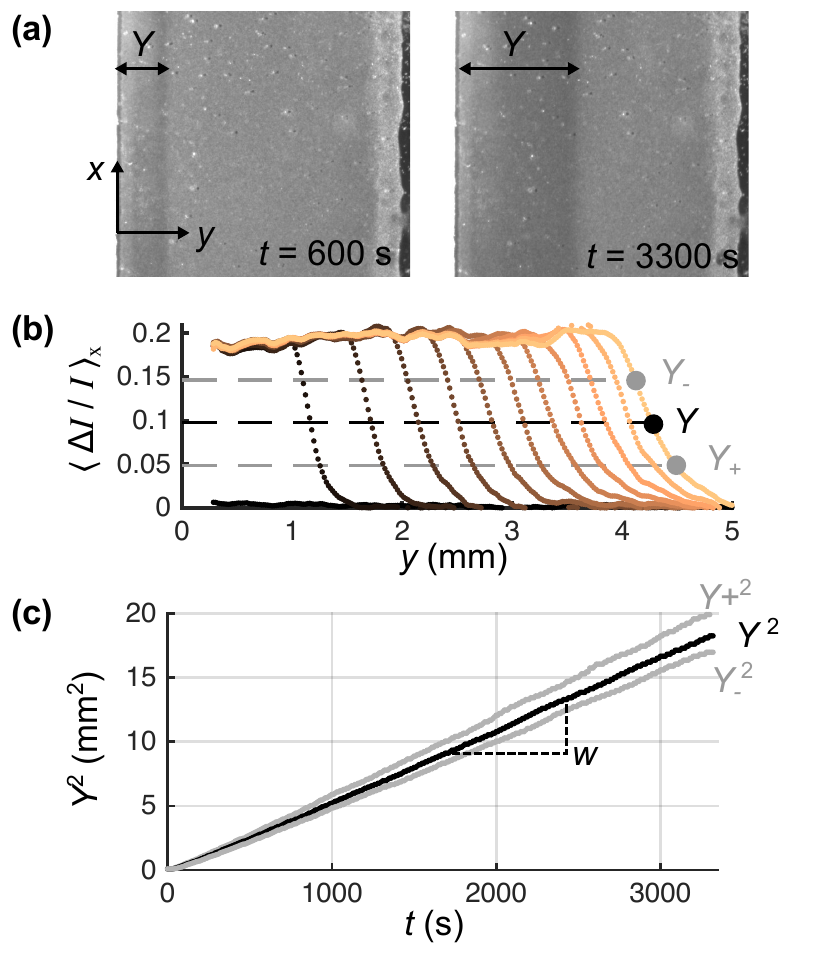}
 	 \caption{
	 \small
	 Imbibition by capillary condensation at high relative humidities ($a>0.6$), here for $a=0.98$. (a) Raw images obtained during the experiment (the field of view is that defined in Figure \ref{fig:Intro}c). (b) Results of image analysis showing the reflectance change as a function of position. Colors represent different times, starting from $t=0$ (black points around $\Delta I / I =0$) and with a time separation of 5 min. The horizontal dashed lines show how the front position is extracted from the data. (c) Front position squared as a function of time.
	  }
  	\label{fig:ImbibitionDynamics}
	\end{center}
\end{figure}

\begin{figure}
  	\begin{center}
	\includegraphics[scale=1]{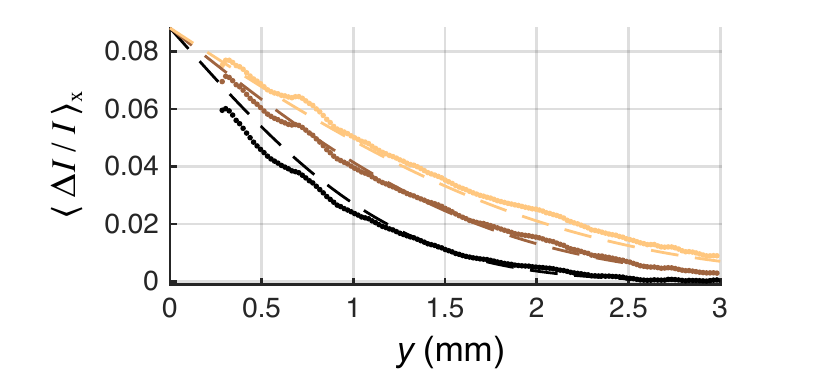}
 	 \caption{
	 \small
	 Diffusion-like dynamics at low relative humidities ($a<0.6$), here for $a=0.50$. Colors represent different times of 40 min, 80 min and 120 min after the beginning of the experiment. The dashed lines are solutions of the diffusion equation, with a constant diffusivity $D=2\times10^{-10} \, \mathrm{m^2/s}$.
	 }
  	\label{fig:DiffusionDynamics}
	\end{center}
\end{figure}

We then recorded dynamics change in reflectance with the sample bonded to glass such as depicted in Figure \ref{fig:Intro}b, during equilibration with a vapor of constant activity $a=p/p_\mathrm{sat}$ after complete evacuation. In such a configuration, exchange of mass was only possible at the open edge (left) of the sample, and we monitored, optically, the resulting lateral propagation of changes in reflectance.

%\Oc{- Note: I am not convinced we should translate the reflectance into $\Psi$ directly, as 1) we argued above that changes in reflectance correspond to changes in adsorbed mass, 2) in dynamic situations, changes in reflectance can come from both local $\Psi$ changes and from local heterogeneities in the advance of the wetting front.}

At high imposed relative humidities ($a>0.6$), the equilibration process resulted in a clearly visible invasion front propagating away from the open edge (Figure \ref{fig:ImbibitionDynamics}a). Image analysis allowed us to extract the front profile as a function of time, after averaging over the $x$ direction. The resulting data (Figure \ref{fig:ImbibitionDynamics}b) showed the propagation of a sharp front separating a fully wet zone (to the left, $\langle \Delta I / I \rangle_x \simeq 0.2$) from a dry zone (to the right, $\langle \Delta I / I \rangle_x = 0$). Note that the fully wet intensity was lower than that on the reflectance isotherm ($0.20$ vs. $0.26$) due to the presence of the glass layer on top of the porous silicon layer. We quantified the front propagation by extracting the mean front position $Y(t)$ and estimated its width using $Y_{+}$ and $Y_{-}$ (see Figure \ref{fig:ImbibitionDynamics}b). As shown in Figure \ref{fig:ImbibitionDynamics}c, the three parameters, $Y$, $Y_{+}$ and $Y_{-}$, exhibited a $\sqrt{t}$ scaling, typical of imbibition dynamics \cite{WickingBook2013,Huber2015}.

For lower imposed vapor pressures ($a<0.6$), the equilibration dynamics was qualitatively different, with a diffusion-like penetration of moisture into the sample rather than a well-defined front (Figure \ref{fig:DiffusionDynamics}). Quantitatively, the dynamics was also much slower, as shown by the very different timescales in Figures \ref{fig:ImbibitionDynamics}b and \ref{fig:DiffusionDynamics}. 
%We discuss the physical origin of the imbibition-like regime and of the diffusion-like regime and their relative dynamics in the next section.

\begin{figure}
  	\begin{center}
	\includegraphics[scale=1]{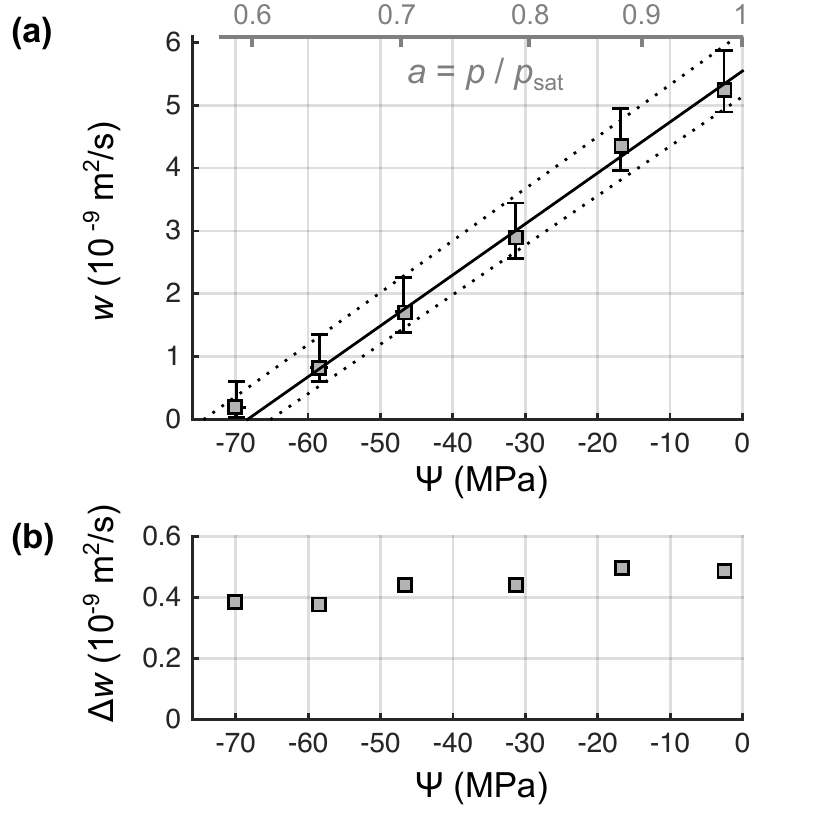}
 	 \caption{
	 \small
	 Front properties for imbibition-like dynamics at high imposed vapor activities ($a>0.6$). (a) Imbibition speed coefficient, $w$, as a function of the Kelvin water potential $\Psi(a)$ (bottom axis) calculated from the activity $a$ (top axis) with Equation \ref{eq:Psi}. Values of $w$ were extracted from experiments such as those in Figure \ref{fig:ImbibitionDynamics}, as the slope of the $Y^2(t)$ data in Figure \ref{fig:ImbibitionDynamics}c. The error bars were obtained from the values $w_{-}$ and $w_{+}$ estimated from $Y_{-}$ and $Y_{+}$, respectively. The lines are linear fits of $w$, $w_{-}$ and $w_{+}$. (b) Imbibition front width, $\Delta w = w_{+} - w_{-}$, as a function of water potential.
	 %\Oc{Add relative humidity axis?}
	  }
  	\label{fig:ImbibitionSpeed}
	\end{center}
\end{figure}

We performed a series of experiments in the imbibition-like regime ($a>0.6$), at different imposed vapor activities, $a$, or equivalently different vapor water potentials $\Psi (a)$ from Equation \ref{eq:Psi}. Using linear fits of the data as in Figure \ref{fig:ImbibitionDynamics}, we extracted the imbibition speed coefficient $w=Y^2/t$ and a typical front width $\Delta w = w_+ - w_-$ with $w_+ = Y_+^2 /t$ and $w_- = Y-+^2 /t$ as a function of $\Psi (a)$. The results are shown in Figure \ref{fig:ImbibitionSpeed} and demonstrate a linear decrease in the imbibition speed as $\Psi$ becomes more negative, as well as a front width that is weakly dependent on the imposed vapor activity.

%In the following section, we discuss the physical origin of the imbibition-like regime and its dynamics relative to the diffusion-like regime.

\section{Discussion}

In all the experiments reported above (Figures \ref{fig:ImbibitionDynamics} \-- \ref{fig:ImbibitionSpeed}), we started with the sample in a dry state ($a=0$), and the equilibration with a vapor at $a>0$ resulted in moisture uptake in the pores. As a result, the processes involved in the dynamics relate to the adsorption branch of the isotherm. From the reflectance isotherm of Figure \ref{fig:Isotherm}, it is clear that the value $a = 0.6$ marked a transition: for $a > 0.6$, capillary condensation resulted in complete filling of all the pores, while for $a < 0.6$ the sample was only partially saturated. Consequently, it is reasonable to assume that in dynamic situations, imposing $a>0.6$ resulted in the  saturation with liquid of the pores at the sample edge, similarly to what would happen if the sample were directly dipped in the bulk liquid, a case classically described with Lucas-Washburn dynamics.
Here, however, the samples were not in contact with bulk liquid but with a subsaturated vapor; in this scenario, we expect that an additional meniscus exists at the open end of the pore, as depicted in Figure \ref{fig:Intro}g. The driving force for the liquid flow thus resulted from the competition of the capillary pressure of two menisci -- one at the open edge and the other at the advancing front (Figure \ref{fig:Intro}g) -- leading to the modified Lucas-Washburn dynamics expressed in Equation \ref{eq:Washburn}, with a natural driving force proportional to $\Psi(a) - \Psi_\mathrm{c}$; $\Psi (a)$ represents the capillary pressure imposed by the local liquid-vapor Kelvin equilibrium (Equation \ref{eq:Kelvin}) at the sample edge, while $\Psi_\mathrm{c}$ represents the intrinsic capillary pressure associated with the pore geometry and wetting properties (Equation \ref{eq:PsiC}). The classical Washburn relation is recovered in the limit $a \rightarrow 1$, i.e. when the vapor approaches saturation; our model thus predicts that imbibition dynamics is identical when using saturated vapor and when using bulk liquid as the boundary condition at the sample edge.

Experimental results as in Figure \ref{fig:ImbibitionSpeed}a show excellent agreement with the linear $w(\Psi)$ dependence predicted by Equation \ref{eq:Washburn}. Furthermore, the $w=0$ intercept and the slope of the linear fits in Figure \ref{fig:ImbibitionSpeed}a allows us to estimate from Equation \ref{eq:Washburn} both the intrinsic capillary pressure $\Psi_\mathrm{c} = -70 \pm 5$ MPa and the Darcy permeability of the porous layer $\kappa = 1.82 \pm 0.05$ $\mathrm{m^2/(Pa.s)}$ independently, using $\phi=0.45$ \cite{Vincent2016}. These values are almost identical to those obtained from drying-induced permeation experiments on similar porous silicon layers that we reported recently \cite{Vincent2016}. This excellent agreement validates our interpretation of the imbibition-like dynamics and the applicability of the modified Lucas-Washburn equation that takes into account the Kelvin capillary pressure (Equation \ref{eq:Washburn}); it also suggests that our assumption that the optical signal, $\Delta I / I$, reflects the local mass uptake in the medium was appropriate.

The success of this modified Lucas-Washburn treatment (Equation \ref{eq:Washburn} applied to Figure \ref{fig:ImbibitionSpeed}a) indicates that imbibition from a saturated vapor ($a=1$, $\Psi = 0$) is equivalent to imbibition from bulk liquid, when local equilibrium at the exposed edge is ensured. This equivalence has been debated in the membrane science community under the name "Schroeder's paradox", based on observation of distinct absorption properties from pure liquid and saturated vapor in swelling media \cite{Choi2003,Vallieres2006}. Neuman and colleagues have provided an explanation of this effect as arising from history-dependent properties of certain materials (e.g. Nafion) \cite{Onishi2007}. For rigid membranes with well-defined pore structure, such as the porous silicon studied here, our observations support the expectation that imbibition dynamics should be defined by the chemical potential at the boundary, independent of the phase that imposes this boundary state.

Classical Lucas-Washburn imbibition experiments in bulk liquids only provide a measure of the product $\kappa \times \Psi_\mathrm{c}$\cite{WickingBook2013,Gruener2009}, except if the pressure $P$ of the liquid reservoir in contact with the sample can be varied: in this latter case, the $w(P)$ response allows the estimation of $\kappa$ and $\Psi_\mathrm{c}$ separately, but requires the use of large positive pressures of magnitude comparable to $\Psi_\mathrm{c}$ \cite{Li2012}. The technique proposed here uses a similar strategy, but the pressure at the edge is brought to large, \emph{negative} values, taking advantage of the fact that small changes in vapor activity result in large changes in the Kelvin stress (Equation \ref{eq:Kelvin}). We recently reported another method exploiting Kelvin stresses to measure both $\kappa$ and $\Psi_\mathrm{c}$, using drying-induced, steady-state permeation flows \cite{Vincent2016}. Globally, using vapor-liquid equilibrium provides a convenient way to vary the capillary stress in the pore liquid (i.e. the driving force for the flow) in a continuous manner through Kelvin equation.

The parameters $\kappa$ and $\Psi_\mathrm{c}$ are associated with the viscous drag in the pores and with capillary effects, respectively, and their measurement offers an interesting way to probe the fluid mechanics of confined liquids.  As we have discussed recently \cite{Vincent2016}, the measured values of $\kappa$ and $\Psi_\mathrm{c}$ suggest that the fluidity and capillarity of water are not noticeably modified in nanoscale confinements: using Laplace law of capillarity (Equation \ref{eq:PsiC}) with $\theta = 25\pm5 \degree$ and the bulk surface tension of water $\sigma = 0.073$ N/m \cite{Vincent2016,Kovacs2009}, we estimate a pore size of $r_\mathrm{p} = 1.9 \pm 0.2$ nm, compatible with pore size estimates using Kelvin-Laplace equation on the reflectance isotherm obtained here ($r_\mathrm{p}=1.8 \pm 0.5$ nm) \cite{Note-2},
or BJH analysis from $N_\mathrm{2}$ mass isotherms ($r_\mathrm{p}=1.4 \pm 0.4$ nm) and dynamic permation flow measurements ($r_\mathrm{p}=1.7\pm0.2$ nm) obtained on a separate sample \cite{Vincent2016}. The pore size value of $r_\mathrm{p}=1.9 \pm 0.2$ nm is also compatible with the Carman-Kozeny relation $\kappa = \phi r_\mathrm{eff}^2 / (8 \eta \tau)$ using the bulk water viscosity $\eta = 1.14 \times 10^{-3} \, \mathrm{Pa.s}$, the tortuosity of the porous layer $\tau = 4.5$, and an effective hydraulic radius $r_\mathrm{eff}=r_\mathrm{p} \left(1 - d/r_\mathrm{p} \right) ^2$ that accounts for the immobility of one monolayer of thickness $d=0.31$ nm at the pore wall \cite{Vincent2016}. This consistency, as well as the compatibility of the observed imbibition dynamics with the stresses predicted from Kelvin equation, provide another demonstration of the excellent extension of the macroscopic laws of capillarity, viscous flow, and thermodynamics at the nanoscale \cite{Gruener2009,Vincent2016}.

%\Oc{Note: applying the hemispherical Kelvin equation to the range $p / p_\mathrm{sat} = 0.45 \-- 0.65$ corresponding to the hysteresis cycle location in the isotherm yields the pore size distribution $r_\mathrm{p} = 1.8 \pm 0.5$ nm.}

According to Equation \ref{eq:Washburn}, when the external vapor activity $a$ is low enough such that $\Psi (a) = \Psi_\mathrm{c}$, there is no net capillary force driving bulk liquid flow. From Equation \ref{eq:Kelvin},  $\Psi (a) = \Psi_\mathrm{c}$ with the value of $\Psi_\mathrm{c}$ determined above corresponds to $a = 0.59 \pm 0.02$, in accordance with the experimentally observed transition to a diffusion-like regime below $a \simeq 0.6$.  In the diffusion-like regime, while there is no capillary driving force for equilibration, there is a chemical potential driving force, as the dry sample needs to uptake some water to come to equilibrium (see Figure \ref{fig:Isotherm}). As a result of this chemical potential imbalance, mass diffusion occurs through the porous medium, explaining the diffusion-like dynamics below $a=0.6$.

 We estimated the effective diffusivity in that regime by fitting the data as in Figure \ref{fig:DiffusionDynamics} with analytical solutions of the one-dimensional diffusion equation with constant diffusivity $D$ \cite{Crank1979}. We note that the diffusivity in Equation \ref{eq:VaporPoroelasticity} is not constant a priori, so the simple fitting with a constant $D$ only provides a way to discuss order of magnitudes. As can be seen in Figure \ref{fig:DiffusionDynamics}, the hypothesis of a constant diffusivity fits the data reasonably well. The diffusivity obtained from the fit was similar ($D \simeq 2 \times 10^{-10} \, \mathrm{m^2/s}$) for two experiments in the diffusion-like regime, one below the hysteresis cycle ($a = 0.35$) and one in the capillary condensation regime ($a = 0.50$). 
 This value may seem low compared to typical bulk molecular diffusivities, but $D$ is in fact not a true molecular diffusivity, but rather incorporates effects from both transport coefficients in the porous medium and local adsorption thermodynamics. Indeed, molecules uptaken in the pores are either adsorbed locally, or transported further in the medium, as illustrated by Equation \ref{eq:VaporDiffusivity}. Based on Knudsen diffusion, Equation \ref{eq:VaporDiffusivity} however predicts $D$ two orders of magnitude smaller than the experimental value (see Theory). This disagreement suggests that surface transport in the adsorbed water at the pore walls may play an important role in the diffusion dynamics. In fact the observed value of $D$ is typical of that reported for surface flows in porous glass and porous silicon \cite{Dvoyashkin2009}.
 
 %\Oc{- I am not sure that we need to discuss diffusion more than this, as we cannot discriminate mechanisms, and we do not have lots of diffusion data. Two additional remarks could be added : 1) The detailed physical processes of multiphase diffusion in nanopores are notoriously complex, including surface diffusion, multilayer diffusion, interaction with geometrical disorder, thermally activated relaxations of metastable states, etc., 2) Equation \ref{eq:VaporDiffusivity} allows to understand why $D$ is the same below the hysteresis and in the hysteresis zone: it's expected that $k$ increases in the hysteresis zone because transport is more efficient (more liquid phase), but the adsorption curve also becomes steeper.}
 
 %\Oc{- I have removed this remark, as I don't think it is reasonable to model diffusion as disjoining-pressure propagation in thin films. We could leave it as a footnote, maybe? -- In fact, we note that a disjoining-pressure based model for the propagation of thin films predicts an effective diffusivity $D=A/(6\pi e)$ where $e$ is the thickness of the adsorbed film and $A$ is the film Hamaker constant \cite{Joanny1986}. Using $A \simeq 10^{-20}$ J \cite{IsraelachviliBook}, $e \simeq 1$ nm and $\eta \simeq 10^{-3}$ Pa.s yields $D \simeq 5 \times 10^{-10} \, \mathrm{m^2/s}$, a value similar to that measured here. }
 %Other authors have proposed mechanisms based on thermally-activated transitions in a landscape of metastable states to explain slow diffusive-like behavior during adsorption \cite{Valiullin2006}. 

Regardless of the physical mechanism behind the effective diffusion process, we note that both the imbibition-like regime and the diffusion-like regime result in invasion dynamics scaling as $\sqrt{t}$, resulting in equilibration times $\tau \sim L^2/w$ and $\tau \sim L^2/D$ respectively, for a sample of dimension $L$. As $w$ and $D$ are separated by more than an order of magnitude, equilibration in the imbibition-like regime is much more efficient than in the diffusion-like regime. Also, since both processes display the same qualitative dynamics $\propto \sqrt{t}$, global mass uptake experiments are not able to distinguish between them, and only spatially resolved measurements as presented here can unveil their contributions.

%\Oc{- Make a comment about the timescale of equilibration during the isotherm measurement (thickness $L=5 \, \mathrm{\mu m}$), predicted to be $0.1$ s in the diffusion regime).?}

Finally, we comment on the width of the imbibition-like front (Figures \ref{fig:ImbibitionDynamics}c and \ref{fig:ImbibitionSpeed}b): since both $Y_+$ and $Y_-$ scaled as $\sqrt{t}$ (Figure \ref{fig:ImbibitionDynamics}c), the widening of the imbibition front $\Delta Y = Y_+ - Y_-$ also followed a $\sqrt{t}$ scaling. The quantity $\Delta w = w_+ - w_- = (Y_+^2-Y_-^2)/t$ is a measure of that widening; as can be seen in Figure \ref{fig:ImbibitionSpeed}b, $\Delta w$ was nearly independent of the imposed water potential and thus also insensitive to the imbibition speed in our experiments. A $t^{1/2}$ scaling for front broadening during an imbibition process is considered anomalous \cite{Alava2004}, but has been reported recently in nanoporous glass \cite{Gruener2012}; these authors suggested that the anomalous broadening was due to uncorrelated menisci movement during imbibition when the pores are sufficiently long compared to the spacing of lateral connections. Due to limited information about the local structure and connectivity of porous silicon $\langle 111 \rangle$ layers, it is unclear if the pores in our study satisfies these geometrical conditions or if the anomalous front broadening originates from other physics.

%\Oc{Note: For independent cylindrical pores of constant radius but with a distribution of diameters across different pores, the front widening should also be in $t^{1/2}$ [ref? easy to show from Washburn], while for independent pores with randomly varying radius, roughening in  $t^{1/4}$, see \cite{Gruener2012}}.

\section{Conclusion}

We have used the optical signature induced by the uptake of water in a model nanoporous medium to study the dynamics of invasion by capillary condensation in the nanopores as a function of the imposed vapor pressure. In a broad ranges of high vapor pressures, invasion occurred as a well defined front reminiscent of Lucas-Washburn imbibition dynamics, with an additional contribution to the capillary force driving the flow that we have shown to be well described by Kelvin equation. At lower vapor pressures, moisture transport transitioned to a slower diffusion-like behavior due to the vanishing of the net capillary driving force. Both processes (imbibition and diffusion) resulted in mass uptake proportional to $\sqrt{t}$, but with qualitatively different spatio-temporal dynamics, as well as order-of-magnitude difference in timescales. Our analysis also suggested that imbibition from saturated vapor was equivalent to imbibition from bulk liquid, due to the insensitivity of the dynamics to the external phase in contact with the pores.

The vapor pressure dependence of the flow in the imbibition regime offers a way to finely tune the imbibition speed through the Kelvin-induced capillary pressure. This control over the dynamics by using an external, easily tunable parameter (the relative humidity of the vapor) is interesting for both fundamental studies of thermodynamics and dynamics in porous media, and for the development of technologies using the response of materials to humidity changes. For fundamental aspects, we have shown that the imbibition-like response allows an independent measurement of both the intrinsic capillary pressure and of the permeability of the porous medium, providing useful information about the capillary and viscous effects at the nanoscale, respectively; globally, our results also demonstrate the validity of Kelvin equation in pores only $\simeq 4$ nm in diameter. For applications in, for example, sensing and membrane processes, our observations and analysis provide a robust basis for the design of materials at the pore-scale to control imbibition dynamics.
%In an applied context, the regimes elucidated here should help interpret and optimize transients in nanopore-based .

%\Oc{- Note again that Kelvin works at the nanoscale?}

%\section{Remarks and Questions}

%\Oc{- It's not obvious why porous silicon gets darker when it's wet. Basic arguments based on indices of refraction suggest that it would be lighter in color (higher reflectance) when filled with water.}

%\section{\Oc{Notes to myself}}

%\Oc{- The steep desorption branch as well as the optical spike suggest pore blocking, governed by the smaller range of pore sizes --> estimate small pore size.}

%\Oc{- Kelvin at the end of condensation --> estimate large pore size (find paper that supports this, maybe coasne?).}

%\Oc{- Front similar to that in pure liquid water (including typical relative intensity change), cf our recent work \cite{Vincent2016}.}

%\Oc{- No dynamic contact angle expected due to the value of the capillary number?}

\begin{acknowledgement}

The authors thank Glenn Swan for technical support and Antoine Robin for help in building the vacuum system. This work was supported by the National Science Foundation (IIP-1500261), the Air Force Office of Scientific Research (FA9550-15-1-0052), the U.S. Department of Agriculture (2015-67021-22844) and the Camille Dreyfus Teacher-Scholar Awards program, and was performed in part at the Cornell NanoScale Facility, a member of the National Nanotechnology Infrastructure Network (National Science Foundation; Grand No. ECCS-15420819).

\end{acknowledgement}

%\begin{suppinfo}
%
%This will usually read something like: ``Experimental procedures and
%characterization data for all new compounds. The class will
%automatically add a sentence pointing to the information on-line:
%
%\end{suppinfo}


%\bibliography{ImbibitionReferences}

\begin{mcitethebibliography}{36}
\bibitem[Not()]{Note-1}
The imbibition mass flux $(\phi \rho_\mathrm{liq} /2) \sqrt{w/t}$ can be
  compared to a typical condensation flux obtained from the kinetic theory of
  gases $p\sqrt{\rho_\mathrm{liq} v_\mathrm{m}/(2 \pi RT)}$, which is much
  greater than the imbibition flux as long as $t \gg 10^{-3}$ seconds, using
  typical values corresponding to our experiments.\relax
\mciteBstWouldAddEndPunctfalse
\mciteSetBstMidEndSepPunct{\mcitedefaultmidpunct}
{}{\mcitedefaultseppunct}\relax
\EndOfBibitem
\end{mcitethebibliography}

\providecommand{\latin}[1]{#1}
\providecommand*\mcitethebibliography{\thebibliography}
\csname @ifundefined\endcsname{endmcitethebibliography}
  {\let\endmcitethebibliography\endthebibliography}{}

\end{document}